\documentclass[12pt]{article} 
\usepackage{epsfig}
\begin{document}
\begin{flushright}
UA-NPPS/05/2005
\end{flushright}
\vspace*{1.5cm}
\begin{center}
{\Large{\bf Transverse momentum parton distributions inspired by a quark 
potential model}}\\
\vspace{2cm}
F.K. Diakonos, G.D. Galanopoulos and X.N. Maintas\\
\smallskip
{\it Department of Physics, University of Athens, GR-15771 Athens, Greece}\\
\end{center}
\vspace{1.5cm}
\begin{abstract}
We derive a nonperturbative transverse momentum distribution for partons 
using a potential model to describe the quark-quark interaction inside the
proton. We use this distribution to calculate the diferential cross-section 
of $\pi^0$-production for intermediate values of transverse momentum in 
$p-p$ collisions at high energies. Assuming a variable string tension constant
for the quark-quark potential we obtain a very good description
of the experimental data at different energies. The corresponding values of the 
mean transverse momentum of the partons are essentially lower than those obtained 
using a Gaussian transverse momentum parton distribution.
\end{abstract}

\section{Introduction}
During the last decades the tests of the perturbative QCD are focused on the experimental [1-15]
and theoretical study [16-26]
of hard processes like 
direct photon and $\pi^0$-production with large transverse momentum in $pp$, $pA$ and $AA$ 
collisions. These processes offer a unique possibility to determine the parton distribution
functions (PDF) inside the proton as well as the parton fragmentation functions (PFF). In 
particular the good understanding of the $pp$ data is the prerequisite for any attempt to
extract new physics, related to the formation of a quark-gluon plasma phase, from the $pA$ 
and $AA$ data. Extensive studies of the $\pi$ (or $\gamma$) production in $pp$ collisions have 
shown that the transverse momentum distribution $g(k_T)$ of the partons inside the proton 
has to be taken into account for a succesfull description of the observed $p_T$ spectrum [20-23].
In all these investigations one assumes 
a Gaussian form for $g(k_T)$. A new nonperturbative parameter is introduced through this 
approach: the mean instrinsic transverse momentum $<k_T>$ of the partons. Although the data
of some experiments concerning hadron production at large $p_T$ \cite{Sivers76,Owens87}
could be explained with relative small $<k_T>$ values ($\approx 0.3-0.5~GeV$), compatible with 
the Heisenberg uncertainty relation for partons inside the proton, there are a number of other
processes leading to a large mean transverse momentum ($<k_T> \approx 1-4~GeV$), depending on $Q^2$, 
for a description of the corresponding experimental data. Such a value of $<k_T>$ is too high 
and cannot be explained as an internal structure of the proton \cite{Zhang02}. However, 
as mentioned by several authors \cite{Feynman78,Owens87} the form of $g(k_T)$ can influence significantly 
the value of $<k_T>$ as well as its $p_T$ dependence. In the present work we derive a transverse
momentum distribution for the partons inside the proton using a potential quark model which has 
succesfully been used to describe the spectra of mesonic \cite{Martin81} and baryonic \cite{Richard81} 
bound states in the past. 
Following \cite{Richard81,Reyes03} we investigate the three-body quantum mechanical bound state problem 
solving numerically the Schr\"{o}dinger equation and obtaining the single particle transverse momentum 
distribution for the consistuent parton. Our main assumption is that intrinsic transverse momentum effects
are not influenced by the Lorenz boost along the beam axis and therefore could be treated within a nonrelativistic
approach. We use the derived distribution to fit experimental data
for the $pp \to \pi_0~+~X$ process. In particular we investigate the measurements for the $p_T$-spectrum 
of the outcoming $\pi^0$ in three experiments performed at different center of mass energies. It turns 
out that a relatively low mean transverse momentum $<k_T>$ ($\approx O(300~MeV)$), compatible with intrinsic
dynamics inside the proton, for the initial partons is sufficient in order to fit perfectly the 
experimental data. A smooth dependence of $<k_T>$ on $p_T$, which within our approach is induced by 
a corresponding variation of the string tension in the quark-quark potential, is required. Our analysis 
shows that the $k_T$ distribution of the consistuent partons can be strongly influenced by three-body effects 
and the form of the confining potential which have to be taken into account in order to describe correctly the 
experimental data concerning the pion production in $pp$ collisions.   
The paper is organized as follows: in Section 2 we present the parton model differential cross section as 
well as the corresponding kinematics for the $\pi^0$-production in $pp$-collisions. In section 3 we 
derive the intrinsic transverse momentum disrtibution for the partons inside the proton using the
quark potential model of \cite{Martin81}. In section 4 we present our numerical results concerning the
description of the data of three different experiments \cite{CP, WA70, PHENIX} as well as the
corresponding dependence $<k_T(p_T)>$. Finally in section 5 we summarize our study and we discuss possible
extensions of the present analysis.

\section{The $pp \to \pi^0~+~X$ cross section}

In the lowest-order perturbative QCD (pQCD), the differential cross section for the direct
pion production in $pp$ collisions is given by:
\begin{eqnarray}
E_{\pi} \frac{d \sigma}{d^3 p}(pp \rightarrow \pi^0 + X)&=&K \sum_{abcd} \int dx_a dx_b f_{a/p}(x_a,Q^2)f_{b/p}(x_b,Q^2)
\nonumber \\
&\times& \frac{d \sigma}{d \hat{t}}(ab \rightarrow cd) \frac{D_{\pi/c}(z_c,Q^2)}{\pi z_c}
\label{eq:eq1}
\end{eqnarray}
where $f_{i/p}$ ($i=a,b$) are the longitudinal parton distribution functions (PDF) for the colliding partons $a$
and $b$ while $D_{\pi/c}$ is the parton fragmentation function (PFF) for the pion. For the scale $Q$
we use the relation $Q^2=\frac{2 \hat{s}\hat{t}\hat{u}}{\hat{s}^2+\hat{t}^2+\hat{u}^2}$ proposed in
\cite{Feynman78} with $\hat{s},~\hat{t},~\hat{u}$ the usual Mandelstam variables. The variable $z_c$ indicates 
the momentum fraction carried by the final hadron. 
The higher order corrections are taken into account by choosing $K \approx 2$ for the corresponding coefficient
in (\ref{eq:eq1}) in the $p_T$ region of interest. It is straightforward to include partonic transverse degrees
of freedom using the following replacement \cite{Owens87} in the PDFs of eq.(\ref{eq:eq1}):
\begin{equation}
dx_i~f_{i/p}(x_i,Q^2) \longrightarrow dx_i d^2 k_{T,i} g(\vec{k}_{T,i}) f_{i/p}(x_i,Q^2)
\label{eq:eq2}
\end{equation}
with $i=a,b$.   
In order to avoid singularities in the differential cross sections describing the partonic subprocesses we 
introduce a regularizing parton mass $m=0.8~GeV$, as in \cite{Feynman78}, in the Mandelstam variables 
occuring in the denominator of the coresponding matrix elements. The explicit formulas of the relevant partonic
cross sections can be found in \cite{Owens87}. Taking into account the transverse degrees of freedom we get the
following expressions for the variables $\hat{s},~\hat{t},~\hat{u}$:
\begin{eqnarray}
\hat{s}&=& s x_a x_b +  \frac{k^2_{T,a} k^2_{T,b}}{s x_a x_b} -2 \vec{k}_{T,a} \cdot \vec{k}_{T,b} \nonumber \\
\hat{t}&=& -(x_a + \frac{k^2_{T,a}}{s x_a}) \frac{p_T \sqrt{s}}{z_c} + \frac{2}{z_c} \vec{k}_{T,a} \cdot \vec{p}_{T}\nonumber \\
\hat{u}&=& -(x_b + \frac{k^2_{T,b}}{s x_b}) \frac{p_T \sqrt{s}}{z_c} + \frac{2}{z_c} \vec{k}_{T,b} \cdot \vec{p}_{T} 
\label{eq:eq3}
\end{eqnarray}
Due to energy-momentum conservation the momentum fraction of the final hadron $z_c$ is given  by:
\begin{equation}
z_c=\frac{(x_a+\frac{k^2_{T,a}}{s x_a}+x_b+\frac{k^2_{T,b}}{s x_b})p_T \sqrt{s}-2 (\vec{k}_{T,a}+\vec{k}_{T,b}) \cdot \vec{p}_T}
{\hat{s}}
\label{eq:eq4}
\end{equation}
For a consistent description of the kinematics in the partonic subprocesses we imply the cuts:
\begin{equation}
z_c \leq 1~~~~;~~~~k^2_{T,a}~<~p_T \sqrt{s}~~~~;~~~~k^2_{T,b}~<~p_T \sqrt{s}
\label{eq:eq5}
\end{equation}
To calculate the cross section given in eq.(\ref{eq:eq1}) we have first to determine the distribution $g(\vec{k}_T)$ and then 
perform the corresponding phase space integrations. Contrary to the usual treatment assuming a Gaussian form for $g(\vec{k}_T)$
we will here derive an alternative expression based on a widely applied quark potential model. 

\section{The intrinsic transverse momentum distribution $g(\vec{k}_T)$}

Since the early days of quantum chromodynamics the main and almost unique tool used for the description of the hadronic bound states 
(mesons, baryons) remain potential models for the quark-quark and quark-antiquark pair interaction. One of the most succesfull models 
was proposed by A. Martin used first to describe mesonic states \cite{Martin81} and later to describe baryons \cite{Richard81}. The 
quark-quark interaction within this model is given as:
\begin{equation}
V(r)=A_{qq} r^{0.1} + B_{qq}
\label{eq:eq6}
\end{equation}
and a similar expression with adapted coefficients $A_{\bar{q}q}$, $B_{\bar{q}q}$ holds for the quark-antiquark interaction. In the
following we will consider a baryon consisting of 3 valence quarks interacting pairwise with the potential (\ref{eq:eq6}).
The Hamiltonian operator of the system is given as:
\begin{equation}
\hat{H}=-\frac{\hbar^2}{2 m} (\nabla^2_1 + \nabla^2_2 + \nabla^2_3) + V(\vec{r}_{12}) + V(\vec{r}_{23}) + V(\vec{r}_{31})
\label{eq:eq7}
\end{equation}
The baryonic ground state can then be obtained by solving the corresponding Schr\"{o}dinger equation. Following \cite{Richard81}
it is convenient to introduce Jacobi coordinates $\vec{\xi}_i$:
\begin{eqnarray}
\vec{\xi}_1&=&\vec{r}_2-\vec{r}_1 \nonumber \\
\vec{\xi}_2&=&\frac{2 \vec{r}_3 - \vec{r}_2 - \vec{r}_1}{\sqrt{3}} \nonumber \\
\vec{\xi}_3&=&\frac{\vec{r}_1 + \vec{r}_2 + \vec{r}_3}{3}
\label{eq:eq8}
\end{eqnarray}
getting the equation:
\begin{equation}
\left[-\frac{\hbar^2}{2 M} \vec{\nabla}^2_{\xi_3}-\frac{\hbar^2}{m}(\vec{\nabla}^2_{\xi_1} + \vec{\nabla}^2_{\xi_2}) + 
\tilde{V}(\vec{\xi}_1,\vec{\xi}_2)\right] 
\Psi_G(\vec{\xi}_1,\vec{\xi}_2,\vec{\xi}_3)=E_G \Psi_G(\vec{\xi}_1,\vec{\xi}_2,\vec{\xi}_3)  
\label{eq:eq9}
\end{equation}
with $M=3 m$ and $m$ is the constituent quark mass. Eliminating the translational mode ($\vec{\xi}_3$) we obtain a partial
differential equation (PDE) depending solely on the variables $\vec{\xi}_1$, $\vec{\xi}_2$. The potential energy $\tilde{V}$
in eq.(\ref{eq:eq9}) is given as:
\begin{equation}
\tilde{V}(\vec{\xi}_1,\vec{\xi}_2)=V(\vec{\xi}_1)+V(\frac{1}{2}(\vec{\xi}_1 - \sqrt{3} \vec{\xi}_2))+
V(\frac{1}{2}(\sqrt{3}\vec{\xi}_2 + \vec{\xi}_1))
\label{eq:eq10}
\end{equation}
Introducing the radial variable $\xi=\sqrt{\xi^2_1+\xi^2_2}$ and the angular variables $\chi,\theta_1,\phi_1,\theta_2,\phi_2$ 
we rewrite eq.(\ref{eq:eq9}) as follows:
\begin{eqnarray}
-\frac{\hbar^2}{m}  \{ \frac{1}{\xi^5} \partial_{\xi} \left( \xi^5 \partial_{\xi} \tilde{\Psi} \right) + \frac{1}{\xi^2}  
\left[ \frac{1}{\sin^2 2 \chi} \partial_{\chi} \left( \sin^2 2 \chi \partial_{\chi} \tilde{\Psi} \right) + 
\frac{\hat{L}^2_1 \tilde{\Psi}}{\cos^2 \chi} + \frac{\hat{L}^2_2 \tilde{\Psi}}{\sin^2 \chi} \right] \} \nonumber \\
+ \tilde{V}(\xi,\chi,\theta_1,\phi_1,\theta_2,\phi_2)\tilde{\Psi}=E_G \tilde{\Psi}
\label{eq:eq11}
\end{eqnarray}
where the angular momentum operators are: 
$$\hat{L}^2_i=-\left[ \frac{1}{\sin \theta_i} \frac{\partial}{\partial \theta_i}
(\sin \theta_i \frac{\partial}{\partial \theta_i})+\frac{1}{\sin^2 \theta_i} \frac{\partial^2}{\partial \phi^2_i} \right]$$ 
The reduced ground state wave function $\tilde{\Psi}$ can be expressed in terms of the hyperspherical harmonics $P_L(\Omega)$
as follows:  
\begin{equation}
\tilde{\Psi}(\xi,\chi,\theta_1,\phi_1,\theta_2,\phi_2)=\sum_{L=0}^{\infty} \frac{u_L(\xi)}{\xi^{5/2}} 
P_L(\chi,\theta_1,\phi_1,\theta_2,\phi_2)
\label{eq:eq12}
\end{equation}
The radial part of the wave function $\tilde{\Psi}$ fullfils the equation \cite{Reyes03}:
\begin{equation}
\frac{d^2 u_L}{d \xi^2}-\frac{15/4 + L(L+4)}{\xi^2} u_L + \frac{m}{\hbar^2} E_G u_L - 
\frac{m}{\hbar^2} \sum_{L'} u_{L'} \tilde{V}_{L L'} = 0
\label{eq:eq13}
\end{equation}
As mentioned in \cite{Richard81} the matrix $\tilde{V}_{L L'}=\int d\Omega P_L^*(\Omega) \tilde{V}(\xi,\Omega) P_{L'}(\Omega)$
for the particular choise of the potential (\ref{eq:eq6}) is diagonal dominant and therefore to a very good approximation the
ground state of the baryonic system is determined by the $L=0$ term in the expansion (\ref{eq:eq12}). Therefore the radial
part of the ground state wavefunction obeys, within the above approximation, the equation:
\begin{equation}
\frac{d^2 u_0}{d \xi^2}-\frac{15}{4 \xi^2} u_0 + \frac{m}{\hbar^2} (E_G - \tilde{V}_{00}) u_0 = 0
\label{eq:eq14}
\end{equation}
where: $\tilde{V}_{00}=A_{00} + B_{00} \xi^{0.1}$ with $A_{00}=\frac{3}{2}A_{qq}$ and $B_{00}=\frac{1}{2} \lambda B_{qq}$.
The constant $\lambda$ is given by: $\lambda=\frac{24}{\pi} \Gamma(1.55) \Gamma(1.5) \Gamma(3.05)$. It is straightforward
to show that the ground state wavefunction in the momentum space (conjugate to the space of $\vec{\xi}_i,~i=1,2,3$) is 
given by:
\begin{equation}
\tilde{\Phi}(k_{\xi}^2)=N \int_0^{\infty} d \xi \xi^{1/2} u_0(\xi) \frac{J_2(k_{\xi} \xi)- k_{\xi} \xi 
J_3(k_{\xi} \xi)}{k_{\xi}^2} 
\label{eq:eq15}
\end{equation}
where $J_n$ is the Bessel function of order $n$ and $N$ is a normalization constant. It is useful to determine the transformation 
of the momenta $\vec{k}_{\xi_i}$ to the cartesian momenta $\vec{k}_i$:
\begin{eqnarray}
\vec{k}_{\xi_1}&=& -\frac{1}{2}(\vec{k}_1-\vec{k}_2) \nonumber \\
\vec{k}_{\xi_2}&=& -\frac{1}{2 \sqrt{3}} (\vec{k}_1 + \vec{k}_2 - 2 \vec{k}_3) \nonumber \\
\vec{k}_{\xi_3}&=& \frac{1}{\sqrt{3}} (\vec{k}_1 + \vec{k}_2 + \vec{k}_3) 
\label{eq:eq16}
\end{eqnarray}
The above expressions simplify in the center of mass frame $\vec{k}_{\xi_3}=\vec{0}$ of the baryonic system where the following
relation is valid:
$$k_{\xi}^2=k_1^2+k_2^2+\vec{k}_1 \cdot \vec{k}_2$$
The two-particle density $\rho(\vec{k}_1,\vec{k}_2)$ in this case is given by:
\begin{equation}
\rho(\vec{k}_1,\vec{k}_2)= \vert \tilde{\Phi}(k_{\xi}^2) \vert^2 = \vert \tilde{\Phi}(k_1^2+k_2^2+\vec{k}_1 \cdot \vec{k}_2) \vert^2
\label{eq:eq17}
\end{equation}
From eq.(\ref{eq:eq17}) we obtain the one-particle transverse momentum density $g(\vec{k}_T)$ as:
\begin{equation}
g(\vec{k}_T)=4 \pi \int_0^{\infty} dk_z \int_{-1}^1 dz \int_0^{\infty} d k_2 k_2^2 \vert \tilde{\Phi}(k_T^2 + k_z^2 + k_2^2 + z k_2 
\sqrt{k_T^2 + k_z^2}) \vert^2
\label{eq:eq18}
\end{equation}
where $z$ is the cosine of the angle between the vectors $\vec{k}=(\vec{k}_T,k_z)$ and $\vec{k}_2$. In fact the function $u_0(\xi)$ can 
be obtained only numerically by solving equation (\ref{eq:eq14}) using the Numerov algorithm. Therefore also the transverse momentum 
distribution (\ref{eq:eq18}) is known only numerically. The integrations in eqs.(\ref{eq:eq15},\ref{eq:eq18}) can be performed 
to a great accuracy (relative error $\approx 10^{-6}$) using a mixture of Gauss-Kronrod quadrature and the VEGAS Monte-Carlo integration
routine \cite{Lepage78}. In Fig.~1 we present the density in transverse momentum space of a parton inside the proton obtained by our approach. 
The values of the constants $A_{qq}$ and $B_{qq}$ are choosen according to \cite{Richard81} in order to fit the size and the binding energy of
the proton. The characteristic structure dominating the form of $g(\vec{k}_T)$ within our model is the second maximum at relative high
transverse momenta. This leads, as we will see in the next section, to a reduction, relative to the Gaussian case, of the mean intrinsic 
transverse momentum of the partons needed to describe the experimental data for the $\pi^0$-production at various energies.  

\begin{figure}[ht]
\centerline{\epsfxsize=3.9in\epsfbox{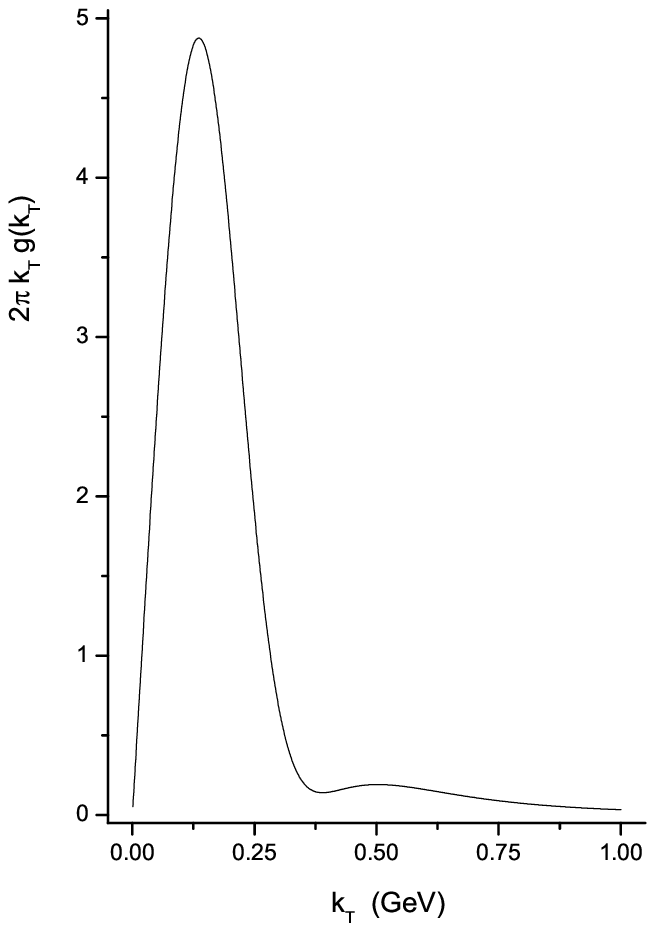}}   
\caption{The intrinsic transverse momentum distribution $2 \pi  k_T g(\vec{k}_T)$ of a parton inside the proton obtained through the 
quark potential model described in section II. \label{fig1}}
\end{figure}

\section{Numerical results}

In the following we will use the derived distribution $g(\vec{k}_T)$ in order to calculate within the parton model the differential
cross section for the $\pi^0$-production in $pp$ collisions according to eq.(\ref{eq:eq1}). For the longitudinal parton distribution
functions we use the recent Martin, Roberts, Stirling and Thorne (MRST) scheme \cite{MRST04} while for the parton fragmentation functions 
we use the Kniehl, Kramer, Potter \cite{Kniehl00} parametrization. The additional nonperturbative parameter in our treatment is the
mean value of the intrinsic transverse momentum $<k_T>$ which is related to the string tension $B_{qq}$ in eq.(\ref{eq:eq6}). Although the
distribution $g(\vec{k}_T)$ is derived for the valence quarks inside the proton here we will use the same distribution also for the
initial gluons assuming universality at the level of consistuent partons. In any case for longitudinal momentum fraction $x_i ~>~0.5$ 
($i=a,b$) the contribution of valence quarks is dominant and our description is accurate.
The phase space integrations are performed using the VEGAS Monte-Carlo
routine. In order to fit the experimental data we allow $<k_T>$ to vary
as a function of the beam energy and the transverse momentum ($p_T$) of the finally produced hadron. We will analyse here the results
of three experiments concerning $\pi^0$-production at different energies. The first set of data are taken from the fixed target experiment
performed in Fermilab (protons incident on $H_2$ target) \cite{CP}. The cross section for the $\pi^0$-production with transverse momentum
$p_T$ at midrapidity and for three different proton beam energies $E_L=200,~300$ and $400~GeV$ is measured. In Fig.~2 we present the various datasets
using symbols while with solid lines we show the results of our calculation and with dashed lines the corresponding results using a Gaussian 
$g(\vec{k}_T)$. As we can see the two descriptions differ only in the region $p_T~<~1~GeV$. Within our approach we need an intrinsic transverse 
momentum $<k_T>$ of the order of at most $0.5~GeV$ in order to reproduce 
all the experimental data. It must be noted that in this case one can perfectly fit the data also for $p_T~<~1~GeV$ which, as we see
in Fig.~2, is not possible using a Gaussian distribution $g(\vec{k}_T)$. In Fig.~3 we show the functions
$<k_T>(p_T)$ obtained using the quark model inspired function $g(\vec{k}_T)$ as well as a Gaussian form. It is evident that
the Gaussian model leads to much higher values of $<k_T>$. In particular this difference is large even for large values of $p_T$
where our approach is more precise as the valence quark contribution is dominant. As already discussed in the previous section 
this difference relies on the fact
that the non-Gaussian $g(\vec{k}_T)$ derived here posseses a second small local maximum at high transverse momenta (see Fig.~1) attributed to the 
form of the quark-quark potential and the many-body character of the system.  

\begin{figure}[ht]
\centerline{\epsfxsize=3.9in\epsfbox{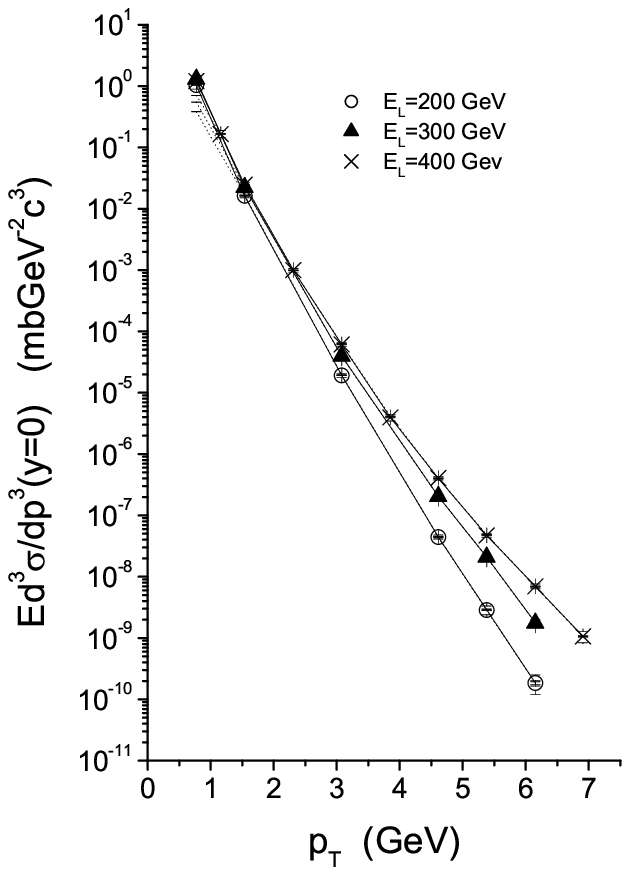}}   
\caption{The differential cross section for the $\pi^0$-production in the Fermilab experiment \cite{CP}. The symbols represent the
experimental datasets for three different beam energies. The solid lines represent the parton model results using a non-Gaussian 
$g(\vec{k}_T)$ while the dotted lines correspond to the analogous results using a Gaussian $k_T$-smearing. Only for $p_T~<~1~GeV$ the
two fits differ. \label{fig2}}
\end{figure}

\begin{figure}[ht]
\centerline{\epsfxsize=3.9in\epsfbox{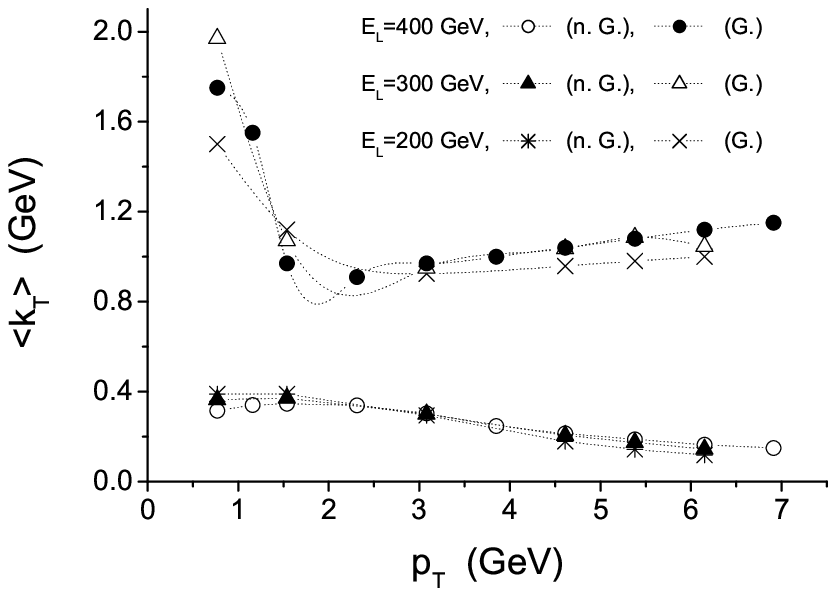}}   
\caption{The functions $<k_T(p_T)>$ for the three datasets presented in Fig.~2. We distinguish between the results of the parton model
using non-Gaussian (n. G.) or Gaussian (G.) intrinsic transverse momentum distribution. 
\label{fig3}}
\end{figure}

The second experiment we consider is the WA70 at CERN SPS \cite{WA70}. It is also a fixed target experiment with $E_L=280~GeV$. We are interested 
in $\pi^0$-production. In Fig.~4 we show the experimental data (full stars) and the corresponding pQCD calculation using the quark model inspired 
$g(\vec{k}_T)$ (solid line) as well as a Gaussian form (dashed line). Both distributions reproduce perfectly the experimental data and cannot be
distinguished graphically. However our approach leads to significantly lower values of $<k_T>$ than
in the Gaussian case. This can be clearly seen in Fig.~5 where we show the function $<k_T(p_T)>$ both for the quark potential model inspired 
$g(\vec{k}_T)$ (full circles) as well as the Gaussian intrinsic transverse momentum distribution (full stars). Also in this
case the difference in $<k_T>$ between the two approaches is large in a $p_T$-region where the main contribution is attributed to the valence 
quarks. For comparison we present also the same function for the Fermilab experiment at $E_L=300~Gev$ (open circles, see Fig.~3). 

\begin{figure}[ht]
\centerline{\epsfxsize=3.9in\epsfbox{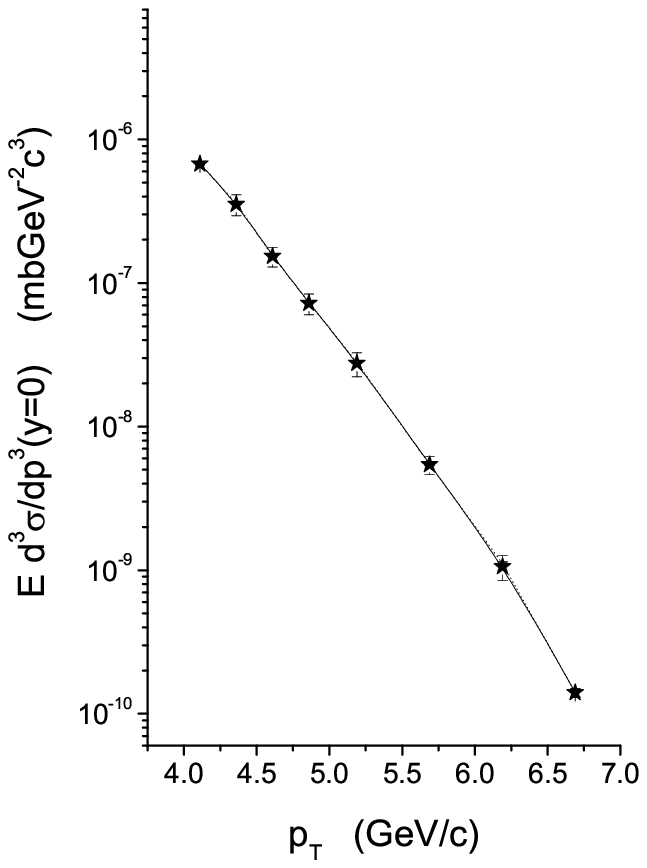}}   
\caption{The differential cross section for $\pi^0$-production in WA70 at $E_L=280~GeV$ (full stars) and the corresponding
parton model calculation using a non-Gaussian (solid line) as well as a Gaussian (dotted line) $g(\vec{k}_T)$. \label{fig4}}
\end{figure}

\begin{figure}[ht]
\centerline{\epsfxsize=3.9in\epsfbox{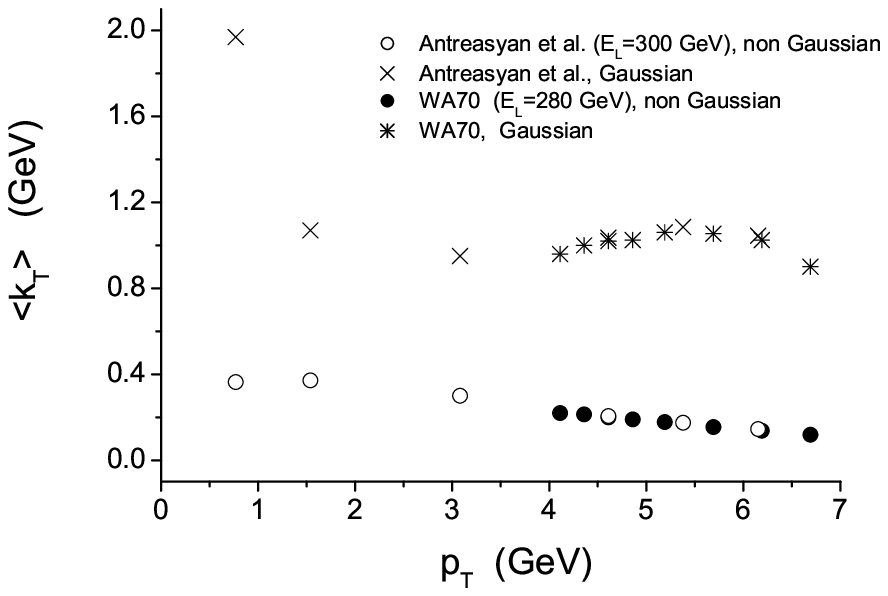}}   
\caption{The dependence $k_T(p_T)$ for the WA70 data set using a non-Gaussian as well as a Gaussian $g(\vec{k}_T)$ in comparison with the corresponding 
behaviour found for the Fermilab experiment at $E_L=300~GeV$ with a non-Gaussian $g(\vec{k}_T)$. \label{fig5}}
\end{figure}
      
Finally we have analysed the $pp \to \pi^0 + X$ data of the most recent PHENIX experiment at the Relativistic Heavy Ion Collider with 
$\sqrt{s}=200~GeV$ \cite{PHENIX}. The corresponding cross section can be described to a good accuracy without inclusion of any $k_T$-smearing,
a fact which is compatible with expectations for the perturbative character of the subprocesses involved in this case. Here we have fitted
the experimental data using non-Gaussian $k_T$-smearing effects. In this way we get a perfect description of the measured cross section. The 
obtained mean intrinsic transverse momentum is almost constant: $<k_T> \approx 250~MeV$. Based on Heisenberg uncertainty relation one could
explain this value of the mean transverse momentum as an effect of the internal partonic structure of the incident proton. Our results are presented 
in Figs.~6,7. The PHENIX data (open cirlces) for $E\frac{d^3 \sigma}{d p^3}$ together with the parton model calculations (crosses) using non-Gaussian 
$g(\vec{k}_T)$ are shown in Fig.~6. The corresponding function $<k_T>(p_T)$ is presented in Fig.~7. 

\begin{figure}[ht]
\centerline{\epsfxsize=3.9in\epsfbox{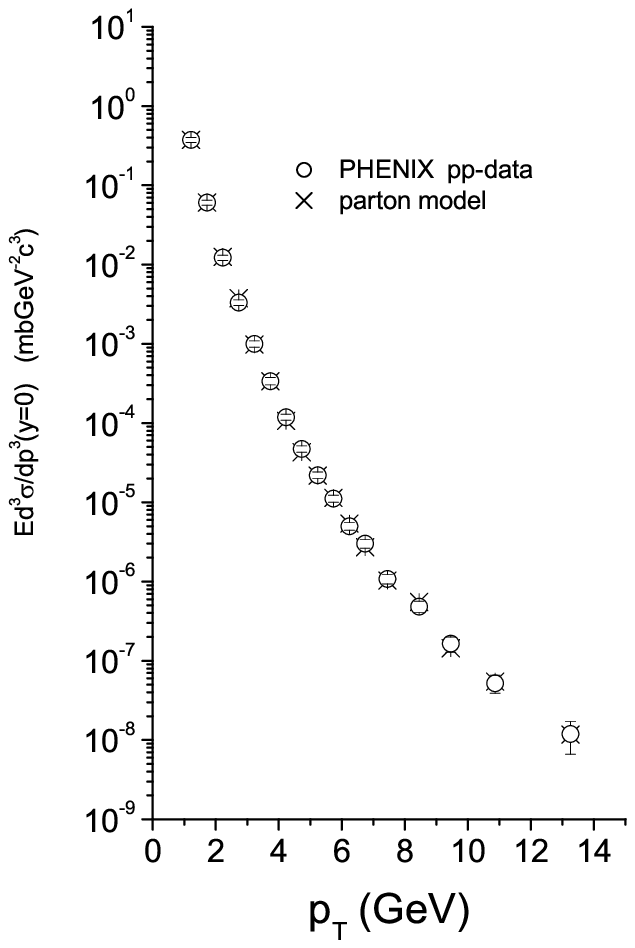}}   
\caption{The $pp \to \pi^0 + X$ differential cross section for the PHENIX experiment ($\sqrt{s}=200~GeV$). The experimental data are presented
by open circles while the crosses indicate the parton model calculation using a non-Gaussian $k_T$-smearing. Only the experimental errors can be seen
at this scale. 
\label{fig6}}
\end{figure}

\begin{figure}[ht]
\centerline{\epsfxsize=3.9in\epsfbox{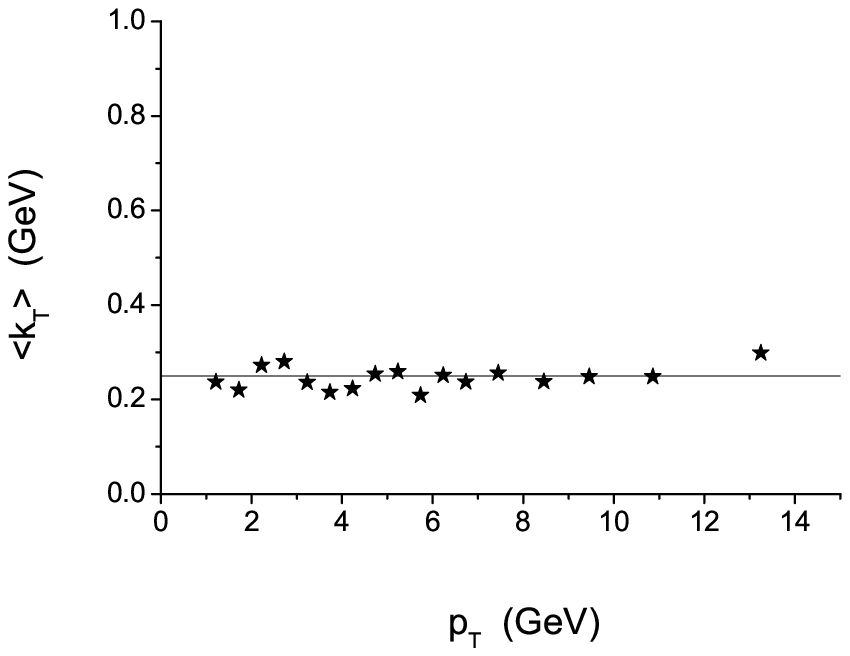}}   
\caption{The function $<k_T(p_T)>$ obtained using the non-Gaussian $g(\vec{k}_T)$ in the parton model description of the PHENIX data. The
solid line at $250~MeV$ is shown as a guide.
\label{fig7}}
\end{figure}
      
\section{Concluding remarks}

Using a quark potential model capable to describe consistently baryonic systems as three quark bound states we have derived an intrinsic transverse
momentum distribution $g(\vec{k}_T)$ of partons inside the proton. This, clearly non-Gaussian, distribution is characterized by the presence of a smooth 
local maximum at relatively high transverse momenta. Our approach is based on the idea that transverse momentum effects may be described nonrelativisticaly
as they are not influenced by the Lorenz boost along the beam axis. Describing the $k_T$-smearing effects in the parton model for $pp$ collisions through 
this non-Gaussian distribution
we calculated the differential cross section for $\pi^0$-production at midrapidity for different beam energies. Assuming that the corresponding
nonperturbative parameter $<k_T>$, related to $g(\vec{k}_T)$, depends on $p_T$ of the finally produced hadron as well as the incident energy, we obtain
a perfect description of the experimental data measured in $pp$ collisions at three different experiments. The corresponding values of $<k_T>$ as a function
of $p_T$ could originate, according to Heisenberg uncertainty relation, from the internal partonic structure of the proton. Our approach is expected to be
valid also in single photon production as well as $pA$ and $AA$ processes \cite{Diak05}. Therefore the performed analysis shows that many-body effects through
a confining potential, reflected at the level of one-particle distributions, may influence strongly the $k_T$-smearing phenomena observed in hadronic 
collisions and therefore should be taken into account for a correct description of the experimental data.  

\vspace*{1.5cm}                                
\noindent
{\bf Acknowledgment} We thank N.G. Antoniou for helpful discussions. This work is financed by EPEAEK in the framework of PYTHAGORAS grants supporting
University research groups under contract 70/3/7420. 

{}

\end{document}